\documentclass[reprint,aps,showpacs,pra,nofootinbib]{revtex4-1}
  \usepackage{color}
  \usepackage{newlfont}
  \usepackage{graphicx}
  \usepackage{amssymb}
  \usepackage{amsmath}
  \usepackage{verbatim}
  \usepackage[latin1]{inputenc}
  \usepackage{bbm}
  \usepackage{multirow}
  \usepackage[thinlines]{easytable}
  \usepackage{braket}
  \usepackage[varg]{txfonts}

\newcommand{\be}{\begin{equation}}
\newcommand{\ee}{\end{equation}}
\newcommand{\beq}{\begin{eqnarray}}
\newcommand{\eeq}{\end{eqnarray}}
\newcommand{\Tr}{\text{Tr}}
\newcommand{\up}{\uparrow}
\newcommand{\down}{\downarrow}
\begin{document}

\title{Nonlocality, quantum correlations, and violations of classical realism in the dynamics of two noninteracting quantum walkers}

\author{Alexandre C. Orthey Jr.}
\author{R. M. Angelo}
\affiliation{Department of Physics, Federal University of Paran\'a, P.O. Box 19044, 81531-980 Curitiba, Paran\'a, Brazil}

\begin{abstract}
That quantum correlations can be generated over time between the spin and the position of a quantum walker is indisputable. The creation of bipartite entanglement has also been reported for two-walker systems. In this scenario, however, since the global state lies in a fourpartite Hilbert space, the question arises as to whether genuine multipartite entanglement may develop in time. Also, since the spatial degrees of freedom can be viewed as a noisy channel for the two-spin part, one may wonder how other nonclassical aspects, such as Bell nonlocality, Einstein-Podolsky-Rosen steering, quantum discord, and symmetrical quantum discord, evolve in time during the walk. The lack of analytical and numerical evidences which would allow one to address these questions is possibly due to the usual computational difficulties associated with the recursive nature of quantum walks. Here, we work around this issue by introducing a simplified Gaussian model which proves to be very accurate within a given domain and powerful for analytical studies. Then, for an instance involving two noninteracting quantum walkers, whose spins start in the singlet state, we quantify the aforementioned nonclassical features as a function of time, and evaluate violations of both realism and related aspects of locality. In addition, we analyze situations in which the initial two-spin state is affected by white noise. The typical scenario found is such that while genuine fourpatite entanglement increases over time, all the investigated nonclassical features vanish (suddenly or asymptotically) except realism-based nonlocality. Moreover, realism is prevented for all finite times. Our findings open perspectives for the understanding of the dynamics of quantum resources in quantum walks.
\end{abstract}

\maketitle

\section{Introduction}

Originally introduced as quantum versions of classical random walks~\cite{aharonov1993quantum}---with some foundational motivations and potential applications to quantum optics---quantum walks have now achieved the status of an ubiquitous tool for studies in areas like quantum computation~\cite{portugal2018quantum,lovett2010universal,childs2013universal}, quantum thermodynamics~\cite{romanelli2014thermodynamics,vallejo2018initial}, and foundations of quantum theory~\cite{robens2015ideal}. Generically speaking, a quantum walk refers to the dynamics of a particle (the walker) whose motion is conditioned to some internal degree of freedom (``the coin''). Some of the usual formulations of this problem consist of confining the walker motion to a dimensionless discrete structure of spacetime and modeling the internal coin with a spin-1/2 algebra. By virtue of the superposition principle, interference patterns typically develop over time, which produces a distinctive mark of quantum walks, namely, ballistic spreading~\cite{ahlbrecht2011}. Interestingly, the mathematical formalism of quantum walks is platform independent, meaning that other physical quantities can be used as internal and external degrees of freedom. In fact, it has been shown that energy levels~\cite{robens2015ideal} or light polarization \cite{schreiber2010photons} perfectly implement the notion of internal coin, whereas the walker position can be suitably emulated with time encoding \cite{schreiber2010photons}, photonic orbital angular momentum~\cite{cardano2015quantum}, or even actual physical position~\cite{crespi2013}. References~\cite{kempe2003quantum,venegas2012quantum} are excellent starting points for the study of quantum walks and Ref.~\cite{wang2013physical} offers a review of physical implementations.

Another relevant feature of a quantum walk is the ability to produce quantum resources. Since information about the spin is shared with the position every time the particle takes a step, quantum correlations are created between these degrees of freedom, especially in the form of entanglement~\cite{carneiro2005entanglement,abal2006quantum}. For instances involving two quantum walkers~\cite{omar2006quantum,peruzzo2010,owens2011,sansoni2012}, the production of nonclassical features becomes even more sophisticated. Different partitions exist and entanglement can be found between the subsystems~\cite{berry2011two,carson2015entanglement}, the spins~\cite{rodriguez2015discord}, and the positions~\cite{goyal2010spatial}. Incidentally, it is precisely the presence of interaction---and entanglement---between the walkers that makes it possible to solve, for example, a wider range of graph isomorphism problems when compared to noninteracting walkers~\cite{gamble2010two}. However, to the best of our knowledge, there is no diagnosis of the presence or dynamical creation of other quantum resources during a quantum walk. Such a resource overview may lead to different perspectives for the use and generation of quantum resources in the fields where the quantum walks apply.

This work aims at advancing the above-delineated framework by dissecting a given two-particle quantum walk with respect to its potentialities in producing several types of quantumness, in particular quantum nonlocality, general quantum correlations, and violations of realism. We analytically assess the behaviors (over time and asymptotically) of some well-established notions, such as entanglement and genuine multipartite entanglement~\cite{horodecki2009ent}, quantum discord~\cite{olliver2001quantum,henderson2001classical}, symmetrical quantum discord~\cite{rulli2011global}, Einstein-Podolsky-Rosen (EPR) steering~\cite{wiseman2007steering,uola2019steering}, Bell nonlocality~\cite{bell1964epr,brunner2014BN}, realism-based nonlocality~\cite{gomes2018nonanomalous}, and irreality~\cite{bilobran2015measure}. While the global state evolves unitarily, thus conserving its initial degree of purity, we find that most of the aforementioned nonclassical features decrease with time between the bipartitions of the system, with some eventual occurences of sudden deaths. On the other hand, some aspects of {\it quantum irreality}---the antithesis of classical realism (full definiteness of all physical quantities)---are shown to persist even when the walkers are arbitrarily far apart from each other and some noise is introduced in the initial two-spin state. This implies that all the involved degrees of freedom remain quantumly linked throughout the time evolution, so that no individual element of reality can be claimed to exist.

This paper is organized as follows. In Section~\ref{sec2}, we present a brief review on quantum walks and introduce a simplified model which proves crucial for our purposes. This model offers considerable analytical power for the treatment of the problem as it avoids the implementation of recursive codes to treat matrices whose dimension increases with time as $16(t+\sigma_0)^2$. In Section~\ref{sec3} we show that genuine fourpartite entanglement increases over time in the global state, thus ``conserving'' the total amount of resource furnished initially. Section~\ref{sec4} provides an exhaustive study of the nonclassical features dynamics associated with the two-spin state, thus regarding the spatial degrees of freedom as an external noisy channel. Concluding remarks are reserved to Section~\ref{sec5}.

\section{Simplified Gaussian model}\label{sec2}

\subsection{One walker}

The state of a one-dimensional quantum walker belongs to a Hilbert space $\mathcal{H}=\mathcal{H}_S\otimes\mathcal{H}_X$, where $\mathcal{H}_S$, spanned by $\{\ket{\up},\ket{\down}\}$, refers to a spin $S=1/2$ space state $(\hbar=1)$ and $\mathcal{H}_X$, spanned by a discrete basis $\{\ket{x}:x\in\mathbb{Z}\}$, denotes the space state associated with the  dimensionless discrete position $X$. Let
\be
\ket{\psi_0}=\left( \cos\tfrac{\alpha}{2}\ket{\up}+\sin\tfrac{\alpha}{2}\ket{\down}\right)\otimes\sum_{x=-\infty}^{\infty} f(x)\ket{x}
\label{psi0}
\ee
be the initial state such that $\alpha\in[0,\pi]$ and $f$ is the initial probability amplitude for the walker position. The single-step unitary evolution is determined by the operator 
\be
U=D\,(C\otimes\mathbbm{1}_X),
\label{U}
\ee 
where $D$ is the conditional displacement operator,
\be
D=\sum_x \Big( \ket{\up}\bra{\up}\otimes\ket{x+1}\bra{x}+\ket{\down}\bra{\down}\otimes\ket{x-1}\bra{x}\Big),
\ee
and $C$ is the so-called quantum coin, a SU(2) matrix which here is chosen to be the Hadamard one:
\be
C\stackrel{\cdot}{=}\tfrac{1}{\sqrt{2}}\begin{pmatrix}
1 & 1 \\ 1 & -1
\end{pmatrix}.
\label{coin}
\ee 
The class of spin states indicated in Eq.~\eqref{psi0} represents a circle in the $xz$ plane of the Bloch sphere, that is, states with no phase difference between $\ket{\up}$ and $\ket{\down}$. Some studies have shown that, when employed along with the Hadamard coin, such class of states is sufficiently general, in the sense that they can yield every possible production rate of spin-position entanglement~\cite{orthey2017asymptotic}, as well as every possible dispersion~\cite{orthey2019connecting}. Still, there is some lack of generality because only one combination of features---entanglement and dispersion---can be simulated through this approach~\cite{orthey2019connecting}.

\begin{figure}[htb]
\centering
\includegraphics[width=\linewidth]{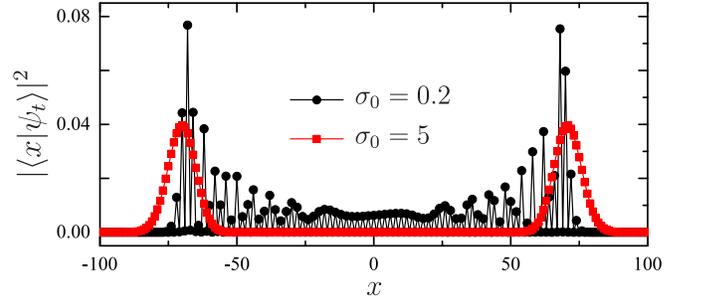}
\caption{Numerical probability distributions $|\bra{x}\psi_t\rangle|^2$ of quantum walks with $\alpha=3\pi/4$ [see Eq.~\eqref{psi0}] for local ($\sigma_0=0.2$) and broad ($\sigma_0=5$) Gaussian states, at $t=100$, as a function of the dimensionless position $x$. The greater the initial dispersion, the more effective the maintenance of the Gaussian shape over time.}
\label{fig1}
\end{figure}

The walker state after $t$ steps can be written as
\be
\ket{\psi_t}=U^t\ket{\psi_0}=\sum_x\Big[a_t(\alpha,x)\ket{\up}+b_t(\alpha,x)\ket{\down}\Big]\otimes\ket{x},
\label{psit}
\ee 
with normalization condition $\sum_x [|a_t(\alpha,x)|^2+|b_t(\alpha,x)|^2]=1$ and a dimensionless time $t\in\mathbb{N}$. If the initial distribution $|f(x)|^2$ is sharply localized, the spin amplitudes $a_t(\alpha,x)$ and $b_t(\alpha,x)$ evolve according to a highly oscillatory pattern, a well-known characteristic of local states (see Fig.~\ref{fig1}). Fourier analysis combined with the stationary phase method~\cite{ambainis2001one} define a largely applied scheme to achieve analytical results, such as those reported for long-time dispersion rates~\cite{brun2003quantum,brun2003pra,ampadu2012brun} and asymptotic entanglement~\cite{abal2006quantum,salimi2012asymptotic,eryiugit2014time,orthey2017asymptotic}. This approach, however, is not appropriate for our purposes because we are interested in looking at the whole dynamics of quantumness quantifiers. To this end, we adopt throughout this work a model according to which the initial distribution is given by the Gaussian function
\be
f(x)=\frac{1}{\sqrt{K}}\exp\text{\small$\left(-\frac{x^2}{4\sigma_0^2} \right)$},\quad\qquad K=\sum_x\exp\text{\small$\left(-\frac{x^2}{2\sigma_0^2}\right)$},
\label{f}
\ee 
where $\sigma_0$ is the dimensionless dispersion and $K$ is the normalization constant. This choice is rather convenient, for it is known that, whenever $\sigma_0$ is sufficiently large, such state preserves not only its Gaussianity over time~\cite{zhang2016creating} but also the interesting properties of ballistic spreading and entanglement creation. Figure~\ref{fig1} gives a comparison of the probability distributions $|\bra{x}\psi_{t}\rangle|^2$ at $t=100$ for quantum walkers initially prepared in a local ($\sigma_0=0.2$) and in a broad ($\sigma_0=5$) Gaussian state.

The region wherein the walker is likely to be found increases with time as $2(t+\sigma_0)$ and the analytical treatment of the problem for long times remains unfeasible even for Gaussian states. We now introduce the fundamental ingredients of our model. First, we employ the approximation
\be
K=\sum_{x=-\infty}^{\infty}\exp\text{\small$ \left(-\dfrac{x^2}{2\sigma_0^2} \right)$}=\vartheta_3(0,e^{-1/2\sigma_0^2})\cong\sqrt{2\pi\sigma_0^2},
\ee
where $\vartheta_3(z,q)=\sum_{x=-\infty}^{\infty}q^{x^2}e^{2kiz}$ is the Jacobi theta function. This approximation fails only for $\sigma_0<1$, a domain that will henceforth be out of scope. Second, for the description of the long-time Gaussian distributions illustrated in Fig.~\ref{fig1}, we propose the ansatz
\begin{subequations}\label{ansatz}
\begin{align}
a_t(\alpha,x)&=q_a^+(\alpha)\,g_t^+(x)+q_a^-(\alpha)\,g_t^-(x),\\
b_t(\alpha,x)&=q_b^+(\alpha)\,g_t^+(x)+q_b^-(\alpha)\,g_t^-(x),
\end{align}
where 
\be
g_t^{\pm}(x)=\text{\small $\frac{(\pm1)^t}{(2\pi\sigma_0^2)^{1/4}}$}\exp\text{\small$\left(-\frac{(x\mp t/\sqrt{2})^2}{4\sigma_0^2} \right)$}
\label{gpm}
\ee 
and 
\be 
q_u^{\pm}(\alpha)=\tfrac{1}{4}\left(\mathfrak{c}_u^\pm\cos\tfrac{\alpha}{2}+\mathfrak{s}_u^\pm\sin\tfrac{\alpha}{2}\right),
\ee 
\end{subequations}
with $\mathfrak{c}_u^\pm$ and $\mathfrak{s}_u^\pm$ $(u=a,b)$ being the coefficients that will adjust our model to the exact numerical results. The above formulas were derived with basis on preliminary numerical studies. Note that the amplitudes $g_t^{\pm}(x)$ move with speed {\small $1/\sqrt{2}$} (a hallmark of the Hadamard walk). The oscillatory form proposed for $q_u^{\pm}(\alpha)$ is naturally induced by the structure of the initial state \eqref{psi0}. After an extensive numerical analysis, involving many different values of $\alpha$, $\sigma_0$, and $t$, we have found
\begin{subequations}\label{csu}
\begin{align} 
\mathfrak{c}_a^\pm&\cong 2\pm \sqrt{2},& \mathfrak{s}_a^\pm&\cong \pm\sqrt{2},\\
\mathfrak{c}_b^\pm&\cong\pm\sqrt{2},& \mathfrak{s}_b^\pm&\cong 2\mp \sqrt{2}.
\end{align}
\end{subequations} 
Equations \eqref{psi0}-\eqref{csu} define our quantum-walk simplified model. The quality of this model was tested via evaluation of the fidelity $|\langle\psi_t^\text{sim}|\psi_t\rangle|^2$ of the state $\ket{\psi_t}$, computed with our simplified model, with respect to the state $\ket{\psi_t^\text{sim}}$, derived via numerical simulation. For sufficiently broad states $(\sigma_0\gtrsim 5)$ and several values of $\{\sigma_0,\alpha,t\}$ the fidelity was never less than 99.8\%. Also noteworthy is the fact that in Eq.~\eqref{gpm} we have tacitly assumed that the dispersion of each Gaussian maintains its initial value $\sigma_0$, which proved to be a rather good approximation whenever $\sigma_0\gg~1$. For future convenience, we note that for $\alpha=0$ and $\alpha=\pi$, which imply the initial states
\begin{subequations}
\begin{align}
&\ket{\up}\otimes\sum_xf(x)\ket{x}\equiv \ket{\psi_0^{\up}}, \\
&\ket{\down}\otimes\sum_xf(x)\ket{x}\equiv \ket{\psi_0^{\down}}, 
\end{align}\label{psi0updown}
\end{subequations}
the above model leads to the respective solutions:
\begin{subequations}
\begin{align}
&\sum_x\Big[a_t(0,x)\ket{\up}+b_t(0,x)\ket{\down}\Big]\otimes\ket{x}\equiv \ket{\psi_t^{\up}}, \\
&\sum_x\Big[a_t(\pi,x)\ket{\up}+b_t(\pi,x)\ket{\down}\Big]\otimes\ket{x}\equiv \ket{\psi_t^{\down}}.
\end{align}\label{psitupdown}
\end{subequations}

Since the global state is pure, the entanglement $E_{SX}(\ket{\psi_t})$ between the spin $S$ and the walker position $X$ can be computed via the linear entropy $\mathcal{L}(\rho_S)=1-\Tr(\rho_S^2)$ of the reduced state 
\be 
\rho_S=\Tr_X\big(\ket{\psi_t}\bra{\psi_t}\big)\stackrel{\cdot}{=}\left(\begin{matrix}\sum_x a_t^2 & \sum_x a_tb_t \\ \\ \sum_x a_tb_t & \sum_x b_t^2 \end{matrix} \right).
\ee 
Simple calculations then show that
\be 
E_{SX}(\ket{\psi_t}):=\mathcal{L}(\rho_S)=\text{\small $\frac{1-\sin{2\alpha}}{4}$}\left(1-\mathcal{E}_t\right),
\label{ESX}
\ee 
where $\mathcal{E}_t=\exp[-t^2/(2\sigma_0^2)]$. We see that maximum (minimum) entanglement production will be attained during the walk when $\alpha=3\pi/4$ ($\alpha=\pi/4$). The factor $\mathcal{E}_t$, which will be ubiquitous in our model, controls the production of entanglement in a way such that the sharper the initial distribution the faster the entanglement production.

\subsection{Two walkers}

Consider now two walkers, named $1$ and $2$, whose state vector lies in the Hilbert space $\mathcal{H}=\mathcal{H}_{S_1}\otimes\mathcal{H}_{S_2}\otimes\mathcal{H}_{X_1}\otimes\mathcal{H}_{X_2}$. In our model, we consider a scenario where the spins $S_{1,2}$ are initially prepared in a maximally entangled state (the singlet state) while the positions $X_{1,2}$ of the walkers are described by Gaussian distributions centered at the origins of their (distinct) coordinate systems. The joint state reads
\be
\ket{\Psi_0}=\text{\small $\frac{\ket{\up\down}-\ket{\down\up}}{\sqrt{2}}$}\otimes\sum_{x_1,x_2} f(x_1)f(x_2)\ket{x_1,x_2},
\label{Psi0}
\ee
with $f$ given by Eq.~\eqref{f}. Moreover, we assume that the walkers do not interact with each other and with the external universe. Concretely, we can conceive an instance such that, after getting their spins correlated, the particles are put to walk in distinct laboratories, which can be arbitrarily separated in space. Each walker is governed by its own unitary dynamics and the eventual emergence of any quantumness between initially independent degrees of freedom ($S_1$ and $X_2$, for example) must be accomplished thanks to the only quantum resource encoded in the joint state, namely, two-qubit entanglement.

To obtain the time-evolved state vector, we should first realize that the state \eqref{Psi0} can be spanned in terms of the kets \eqref{psi0updown} as $\ket{\Psi_0}=\tfrac{1}{\sqrt{2}}(\ket{\psi_0^{\up}}\ket{\psi_0^{\down}}-\ket{\psi_0^{\down}}\ket{\psi_0^{\up}})$. Linearity immediately allows us to use the solutions \eqref{psitupdown} to write 
\be
\ket{\Psi_t}=\tfrac{1}{\sqrt{2}}\left( \ket{\psi_t^\up}\ket{\psi_t^\down}-\ket{\psi_t^\down}\ket{\psi_t^\up} \right),
\ee
which can be more explicitly written as 
\be 
\ket{\Psi_t}=\text{\small $\sum_{x_1,x_2}\frac{\exp\left(-\frac{2t^2+4x_{cm}^2+x_r^2}{8\sigma_0^2}\right)}{\sqrt{2\pi\sigma_0^2}}$}\ket{s_t(x_r)}\otimes\ket{x_1,x_2},
\label{Psit}
\ee 
where we have introduced, for the sake of notational simplicity, $x_{cm}=(x_1+x_2)/2$ (the center of mass position), $x_r=x_2-x_1$ (the relative position), the nonnormalized state
\be
\ket{s_t(x_r)}=\sinh\left(\tfrac{tx_r}{2\sqrt{2}\,\sigma_0^2}\right)\ket{\beta_{23}}+\cosh\left(\tfrac{tx_r}{2\sqrt{2}\,\sigma_0^2}\right)\ket{B_4},
\label{st}
\ee
and the Bell basis
\be
\begin{array}{lll}
\displaystyle \ket{B_1}=\text{\small $\frac{\ket{\up\up}+\ket{\down\down}}{\sqrt{2}}$},& \quad & 
\displaystyle \ket{B_2}=\text{\small $\frac{\ket{\up\up}-\ket{\down\down}}{\sqrt{2}}$},\\ \\
\displaystyle \ket{B_3}=\text{\small $\frac{\ket{\up\down}+\ket{\down\up}}{\sqrt{2}}$}, & &
\displaystyle \ket{B_4}=\text{\small $\frac{\ket{\up\down}-\ket{\down\up}}{\sqrt{2}}$},
\end{array}\label{Bell-states}
\ee
which allowed us to write $\ket{\beta_{23}}\equiv(\ket{B_2}-\ket{B_3})/\text{\small $\sqrt{2}$}$.

It is clear from the result \eqref{Psit} that none of the original degrees of freedom $\{S_1,S_2,X_1,X_2\}$ factorizes for $t>0$. On the other hand, the two-spin state \eqref{st} depends only on the relative coordinate, so that the state associated with the center of mass must factorize. This can be proved as follows. Let us replace laboratory positions $\{x_1,x_2\}$ with center of mass and relative coordinates $\{x_{cm},x_r\}$ through the usual map $\ket{x_1}\otimes\ket{x_2}\mapsto\ket{x_2-x_1}_r\otimes\ket{\tfrac{x_1+x_2}{2}}_{cm}$~\cite{angelo2011,angelo2012}, which links every state in $\mathcal{H}_1\otimes\mathcal{H}_2$ with a counterpart in $\mathcal{H}_{cm}\otimes\mathcal{H}_r$ (assuming walkers with equal masses). Using this map and changing dummy variables in summations, we rewrite the state \eqref{Psit} in an explicitly separable form, $\ket{\Psi_t}=\ket{\Theta}_{cm}\otimes\ket{\Phi_t}_r$, where
\begin{subequations}
\begin{align}
\ket{\Theta}_{cm}&=\sum_{x_{cm}}\text{\small $\frac{\exp\left(-\frac{x_{cm}^2}{2\sigma_0^2} \right)}{(\pi\sigma_0^2)^{1/4}}$}\ket{x_{cm}}, \\
\ket{\Phi_t}_r&=\sum_{x_r}\text{\small $\frac{\exp\left(-\frac{2t^2+x_r^2}{8\sigma_0^2}\right)}{(4\pi\sigma_0^2)^{1/4}}$}\ket{s_t(x_r)}\otimes\ket{x_r}.
\end{align}
\end{subequations}
This completes the proof. An interesting observation can now be made for the spins. By separating the summation for $x_r$ in parcels with $x_r<0$, $x_r=0$, and $x_r>0$, we can compute the asymptotic state $\ket{\Phi_{\infty}}=(\ket{\phi_{\infty}^+}\otimes\ket{S_+}-\ket{\phi_{\infty}^-}\otimes\ket{S_-})/\text{\small $\sqrt{2}$}$, where
\be
\ket{\phi_t^{\pm}}=\sum_{x_r>0}\text{\small $\frac{\exp\left(-\frac{(t-x_r/\sqrt{2})^2}{4\sigma_0^2}\right)}{(4\pi\sigma_0^2)^{1/4}}$}\ket{\pm x_r}, \,\,\,\,
\ket{S_{\pm}}=\text{\small $\frac{\ket{\beta_{23}}\pm\ket{B_4}}{\sqrt{2}}$}, \,\,\,
\ee
and $\phi_{\infty}^{\pm}=\lim_{t\to\infty}\ket{\phi_t^{\pm}}$. We see, therefore, that by measuring the sign of the relative coordinate, one makes the two-spin state collapse to either $\ket{S_+}$ or $\ket{S_-}$, which constitute peculiar coherent superpositions of Bell states. 

\begin{figure}[ht]
\centering
\includegraphics[width=\linewidth]{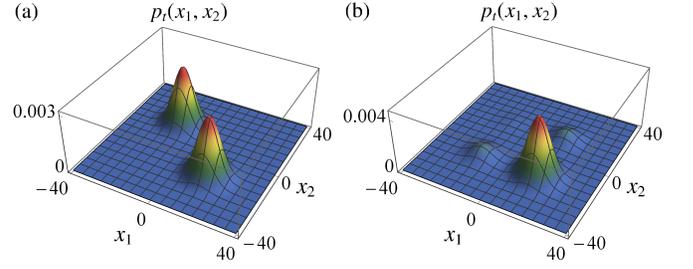}
\caption{Probability distribution $p_{t}(x_1,x_2)$ at $t=20$ of two quantum walkers with initial states (a) $\ket{\Psi_0}$ [see Eq.~\eqref{Psi0}] and (b) $\ket{\psi_0^{\up}}\ket{\psi_0^{\down}}$ [see Eqs.~\eqref{psi0updown}], both with $\sigma_0=5$. A correlated two-spin state is more effective in producing strong spatial anticorrelations.}
\label{fig2}
\end{figure}

Now we show that correlations develop between walkers' positions. The probability $p_t(x_1,x_2)\!=\!\Tr\left(\Omega_t\ket{x_1,x_2}\bra{x_1,x_2}\right)$ of finding them at the respective locations $(x_1,x_2)$, at time $t$, given the state $\Omega_t=\ket{\Psi_t}\bra{\Psi_t}$, results in
\be 
p_t(x_1,x_2)=\text{\small $\frac{e^{-\frac{x_{cm}^2}{\sigma_0^2}}}{4\pi\sigma_0^2}$}\left(\text{\small $e^{-\tfrac{\left(t-x_r/\sqrt{2}\right)^2}{2\sigma_0^2}}+e^{-\tfrac{\left(t+x_r/\sqrt{2}\right)^2}{2\sigma_0^2}}$}\right),
\ee 
whose maximum value occurs for $x_{cm}=0$ and $x_r=\pm\,t\text{\small $\sqrt{2}$}$, that is, $x_1=-x_2=\pm t/\text{\small $\sqrt{2}$}$. This implies a notorious spatial anticorrelation for walkers' positions. Such effect is not observed, for instance, when the joint state is given by $\ket{\psi_t^{\up}}\ket{\psi_t^{\down}}$ [see Eqs.~\eqref{psitupdown}]---a scenario where the walkers start in a fully uncorrelated state and evolve without any interaction. We immediately conclude, therefore, that it is the presence of the initial correlations between the spins that induces the development of spatial correlations (similar results have been reported for local states~\cite{omar2006quantum}). Figure~\ref{fig2} illustrates this result. While the walkers are more likely to be found at the anticorrelated locations $x_1=-x_2=\pm 20/\text{\small $\sqrt{2}$}$ at the instant $t=20$ when the spins start in the singlet state [Fig.~\ref{fig2}(a)], such strong correlation does not appear when the spins are prepared in $\ket{\up\down}$ [Fig.~\ref{fig2}(b)]. Even though the spacetime is modeled as discrete, numerical simulations are throughout presented with continuous variables, which render the results easier to appreciate.

The reliability of our Gaussian model was again checked via direct comparisons with simulations. To this end we evaluate the fidelity $|\braket{\Psi_t^{\text{sim}}|\Psi_t}|^2$ between the state $\ket{\Psi_t^{\text{sim}}}$, computed via numerical simulations, and the state $\ket{\Psi_t}$, derived with our Gaussian model. Some typical results are presented in Table~\ref{tab:fidelity}. We see that the Gaussian model is fairly good for sufficiently broad states ($\sigma_0\gtrsim 3$) and performs better for small times, since in this regime the spreading of the wave packets (not implemented in our model) is less significant.  Similar behaviors for the fidelity were observed for generic spin states, thus indicating the broad adequacy of our model.
\begin{table}[htb]
\centering
  \begin{tabular}{c c c c c c c}
  \hline
  $\sigma_0$ & \qquad & 1 & 2 & 3 & 5 & 10 \\
  \hline\hline
  $t=50$  &   & 0.3284 & 0.9098 & 0.9802 & 0.9947 & 0.9987 \\
  $t=100$ &   & 0.1709 & 0.7973 & 0.9641 & 0.9939 & 0.9987 \\
  \hline
  \end{tabular}
\caption{Fidelity $|\braket{\Psi_t^{\text{sim}}|\Psi_t}|^2$ of the state $\ket{\Psi_t}$, derived through our simplified Gaussian model, with the numerical simulation $\ket{\Psi_t^{\text{sim}}}$, for two noninteracting quantum walkers, at times $t=50$ and $100$, for different values of $\sigma_0$ and the initial state \eqref{Psi0}.}
\label{tab:fidelity}
\end{table}

In possession of solution \eqref{Psit}, we are ready to conduct a thorough study of several quantumness that develop over time in the two-body quantum walk under scrutiny. Basically, we divide the presentation in two parts. In the first, we show that genuine fourpartite entanglement is monotonically generated during the walk. In the second, we consider a Bell scenario where the spatial degrees of freedom constitute noisy channels for the spins and then investigate the time evolution of several quantumness quantifiers.

\section{Genuine fourpartite entanglement}
\label{sec3}

We have seen above that the initial entanglement between the spins induces the development of (presumably quantum) spatial correlations, after all, the walkers do not interact with each other. Naturally, one may ask whether entanglement can also be created among some other degrees of freedom, as for instance between $X_{1(2)}$ and $S_{2(1)}$, or even among all degrees of freedom in an inextricable way. Now we show that the latter type of entanglement does indeed take place.

Our analysis is based on the measure of genuine multipartite entanglement (GME) introduced by Ma {\it et al}. in Ref.~\cite{ma2011measure}. This quantifier is very convenient to our purposes because it assumes a simple computational form for multipartite pure states. Given a pure state $\ket{\Phi}\in\bigotimes_{i=1}^n\mathcal{H}_i$, the authors defined the GME concurrence of $\ket{\Phi}$ as
\be
C_{\text{\tiny GME}}\left(\ket{\Phi}\right):=\min_{\gamma_i\in\gamma}\sqrt{2\,\mathcal{L}(\rho_{\gamma_i})},\label{CGME}
\ee 
where $\mathcal{L}(\rho)$ is the linear entropy of $\rho$ and $\gamma=\{\gamma_i\}$ is the set of all possible parts defining the bipartitions of the state. According to this definition, GME will be present only if the state is nonseparable in every bipartition, that is, if the reduced states $\rho_{\gamma_i}$ of $\ket{\Phi}$ are all nonpure. In our system, two examples of parts $\gamma_i$ are $X_1$ (for the bipartition $X_1|X_2S_1S_2$) and $X_2S_1$ (for $X_2S_1|X_1S_2$), for which one finds the respective reduced states $\rho_{X_1}=\Tr_{X_2S_1S_2}\Omega_t$ and $\rho_{X_2S_1}=\Tr_{X_1S_2}\Omega_t$, with $\Omega_t=\ket{\Psi_t}\bra{\Psi_t}$. Using the state \eqref{Psit} we computed all possible reduced states\footnote{For all practical purposes, the summations over positions, which emerge in the partial trace, can be safely substituted by integrals. This has been checked for Gaussian states with $\sigma_0=5$, in which case the difference between a discrete sum over a closed two-dimensional box of width $200+2t$ and an integral over the whole $\mathbb{R}^2$ was never greater than $10^{-15}$.} $\rho_{\gamma_i}$. For instance, for the two-spin state we found
\be 
\rho_S=\text{\small $\frac{1-\mathcal{E}_t}{2}$}\ket{\beta_{23}}\bra{\beta_{23}}+\text{\small $\frac{1+\mathcal{E}_t}{2}$}\ket{B_4}\bra{B_4},
\label{operator_rho_S}
\ee 
with $S=S_1S_2$. Analytical expressions were then obtained for the respective linear entropies, the results being
\begin{subequations}
\begin{align}
&\mathcal{L}(\rho_{S_j})=\tfrac{1}{2}, & &\mathcal{L}(\rho_{X_j})=\tfrac{1}{2}\left(1-\mathcal{E}_t\right), \\
&\mathcal{L}(\rho_{S_jX_k})=\tfrac{1}{2}, & &\mathcal{L}(\rho_S)=\mathcal{L}(\rho_X)=\tfrac{1}{2}\left(1-\mathcal{E}_t^2\right).
\end{align}\label{L}
\end{subequations}
with $j,k\in\{1,2\}$ and $X=X_1X_2$. From these relations and the definition~\eqref{CGME}, one finds
\be
C_{\text{\tiny GME}}(\ket{\Psi_t})=\sqrt{1-\mathcal{E}_t},
\label{4Et}
\ee
which is a monotonically increasing function of time and can also be written as monotonic functions of the bipartite entanglement quantifiers $\mathcal{L}(\rho_{X_j})$ and $\mathcal{L}(\rho_{X})$. These results indicate that even though each walker evolves independently, the presence of entanglement between the spins at the beginning of the walk allows the global state of the walkers to develop genuine fourpartite entanglement over time. Later on, this interpretation will be corroborated by further evidences. Finally, note that all bipartitions will be equally entangled as $t\to\infty$.

\section{Quantumness dynamics between spins}
\label{sec4}

In this section, we confine our attention to the spins only. This leads us to special Bell scenarios where information about the spins of the particles are encoded, via quantum correlations, on spatial degrees of freedom---a mechanism that tends to degrade the resources present in the two-spin state. As a materialization of such  scenarios, we can envisage instances similar to those recently proposed for witnessing aspects of quantum gravity~\cite{bose2017,marletto2017}, where the spin value defines the path to be taken by the particle (as also happens in a Stern-Gerlach experiment) and then each specific path couples with the gravitation source in a particular manner. In this framework, the spatial degrees of freedom are expected to play the role of a noisy channel, whose effect over the two-spin state varies during the motion of the walkers. We now investigate how several nonclassical features present in the two-state spin vary with time under the aforementioned noisy channel.

To give more generality to our study, we consider that the spins are initially prepared in the Werner state
\be
\rho_{\epsilon}^W=(1-\epsilon)\tfrac{\mathbbm{1}}{4}+\epsilon\ket{B_4}\bra{B_4},
\ee 
with $\epsilon\in[0,1]$. This formulation considers a white noise of amplitude $1-\epsilon$ over the singlet state $\ket{B_4}$. Assuming Gaussian amplitudes for the positions, as in Eq.~\eqref{Psi0}, the initial state of the two-walker model becomes $\rho_0=\rho_{\epsilon}^W\otimes \ket{\varphi_1,\varphi_2}\bra{\varphi_1,\varphi_2}$, where $\ket{\varphi_i}=\sum_{x_i}f(x_i)\ket{x_i}$. Applying the time evolution operator $U_1^tU_2^t$ [see Eq.~\eqref{U}] and tracing over the positions yield
\be
\rho_t^{\epsilon}=(1-\epsilon)\tfrac{\mathbbm{1}}{4}+\epsilon\,\rho_S,
\label{rho_werner}
\ee
with $\rho_S$ being the time-dependent density operator~\eqref{operator_rho_S}. From now on, we restore the time dependence in the notation. The purity of the two-spin state $\rho_t^{\epsilon}$ reads
\be
\mathcal{P}(\rho_t^{\epsilon})=\Tr\left[(\rho_t^{\epsilon})^2\right]=\tfrac{1}{4}\left[1+\epsilon^2\left(1+2\mathcal{E}_t^2 \right) \right],
\ee
which monotonically decreases with time (and with the noise amplitude $1-\epsilon$) towards the asymptotic value $(1+\epsilon^2)/4$. This shows that the spatial variables indeed get more correlated with the spins as the walk takes place. Moreover, the decoherence and the whole dynamics of the two-spin state are controlled by the decay factor $\mathcal{E}_t$ which, by its turn, is determined by the initial dispersion $\sigma_0$ of the Gaussian amplitudes. The broader the spatial distributions, the larger the time scale within which the two-spin state keeps its coherence. Accordingly, a completely delocalized walker $(\sigma_0\to\infty)$ will never have its position correlated with its spin during the walk. This is reasonable since, in this case, it is difficult to defend that, being everywhere, the walker really walks.

\subsection{Bell nonlocality}

Our analysis starts by considering Bell nonlocality~\cite{bell1964epr,brunner2014BN}. A quantum state is termed Bell nonlocal if its underlying probability distributions do not admit a local hidden variable model. In practice, this is signalized by violations of Bell inequalities. One of particular convenience is the Clauser-Horne-Shimony-Holt (CHSH) inequality~\cite{chsh1969proposed}
\be
\mathbb{B}_t=\big|\braket{A_+ B_+}+\braket{A_- B_+}+\braket{A_+ B_-}-\braket{A_- B_-} \big| \leq 2,
\label{Bchsh}
\ee 
where $\braket{A_jB_k} = \Tr[ \rho_t^{\epsilon} A_j\otimes B_k]$. Here, $A_{\pm}$ and $B_{\pm}$ denote observables acting on $\mathcal{H}_{S_1}$ and $\mathcal{H}_{S_2}$, respectively. 

\begin{figure}[htb]
\centering
\includegraphics[width=\linewidth]{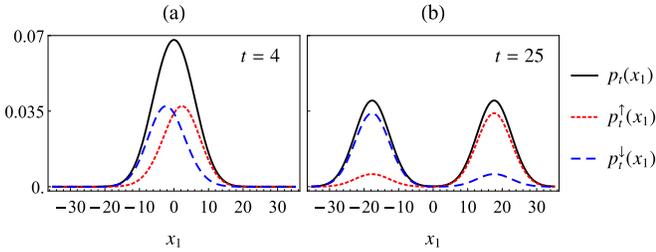}
\caption{Probability distributions $p_t(x_1)$ (black line), $p_t^\up(x_1)$ (dotted red line), and $p_t^\down(x_1)$ (dashed blue line) of finding particle $1$ at position $x_1$, at position $x_1$ with spin up, and at position $x_1$ with spin down, respectively, at instants (a) $t=4$ and (b) $t=25$. The initial dispersion is $\sigma_0=5$.}
\label{fig3}
\end{figure}

As a first step, it is instructive to look at the nonlocality induced by $\rho_t^{\epsilon}$ when there is no white noise ($\epsilon=1$ and $\rho_t^{\epsilon=1}=\rho_S$). It can be directly demonstrate by taking $A_{\pm}$ and $B_{\pm}$ in the form $\hat{v}_i\cdot\vec{\sigma}$, with $\vec{\sigma}=(\sigma_1,\sigma_2,\sigma_3)$ being the vector composed of Pauli matrices  and $\hat{v}_i\in\mathbb{R}^3$ unit vectors. By letting $\hat{v}_i$ assume orthogonal directions $\hat{e}_1$ and $\hat{e}_2$, for particle $1$, and $-(\hat{e}_1+\hat{e}_2)/\text{\small$\sqrt{2}$}$ and $(-\hat{e}_1+\hat{e}_2)/\text{\small$\sqrt{2}$}$, for particle $2$, one shows that $\mathbb{B}_t=\left(1+3\mathcal{E}_t\right)/\text{\small$\sqrt{2}$}$, which implies a CHSH-inequality violation for
\be
t<\sigma_0\text{\small $\sqrt{2\ln\left(\dfrac{3}{2\sqrt{2}-1} \right)}$}\cong 0.995\,\sigma_0.
\label{violation}
\ee
This means that for $t>\sigma_0$ Bell nonlocality will no longer be detected with those specific measurement directions. This can be explained as follows. For long times, each walker's spatial distribution gets sufficiently correlated with its spin. This effect is illustrated in Fig.~\ref{fig3}, where the probability distributions $p_t(x_1)=\Tr(\Omega_t \ket{x_1}\bra{x_1})$ and $p_t^{\mu}(x_1)=\Tr(\Omega_t\ket{x_1}\bra{x_1}\otimes\ket{\mu}\bra{\mu})$ ($\mu=\,\up,\down$) for the particle $1$ (similarly for particle $2$) are shown at two different instants: (a) just before the Bell-nonlocality sudden death and (b) long after this. As a consequence of the correlations generated between spin and position (of each walker), the power of the noisy channel on the two-spin state increases and the nonlocal correlations degrade. 
 
Instead of simply diagnosing the presence of Bell nonlocality, we now want to quantify it. To this end, we adopt a usual strategy according to which one takes the maximal violation of the inequality~\eqref{Bchsh} as a quantifier for the degree of nonlocality of the state. Here we follow the approach put forward in Refs.~\cite{costa2016quantification,costa2016generalized}. We start with Luo's result~\cite{luo2008quantum}, which ensures that every two-qubit state can be written, up to local unitary operations, as
\be
\zeta=\tfrac{1}{4}\left(\mathbbm{1}\otimes\mathbbm{1}+\vec{a}\cdot\vec{\sigma}\otimes\mathbbm{1}+\mathbbm{1}\otimes\vec{b}\cdot\vec{\sigma}+\sum_{i=1}^{3}c_i\,\sigma_i\otimes\sigma_i \right).
\label{zeta}
\ee
where $\{\vec{a},\vec{b},\vec{c}\}\in\mathbb{R}^3$. For this state, the Bell-nonlocality quantifier proposed in Ref.~\cite{costa2016quantification} can be expressed in the form
\be
\mathcal{B}(\zeta)=\max\left\{\text{\small$ 0,\frac{\sqrt{\vec{c}\cdot\vec{c}-c_{\min}^2}-1}{\sqrt{2}-1}$} \right\},
\ee
where $c_{\min}=\min\{|c_1|,|c_2|,|c_3|\}$. Adapted to the form~\eqref{zeta}, the state~\eqref{rho_werner} is such that $\vec{a}=\vec{b}=\vec{0}$ and $\vec{c}=(-\epsilon,-\epsilon\,\mathcal{E}_t,-\epsilon\,\mathcal{E}_t)$, from which we find
\be
\mathcal{B}(\rho_t^{\epsilon})=\left(1+\sqrt{2}\right)\max\left\{0,\epsilon\sqrt{1+\mathcal{E}_t^2}-1\right\}.
\label{Bt}
\ee 
It follows that Bell nonlocality will be present only for 
\be
t<\sigma_0\text{\small $\sqrt{\ln\left(\frac{\epsilon^2}{1-\epsilon^2}\right)}$}\equiv t_\mathcal{B} .
\label{tB}
\ee 
This shows that for any state $\rho_t^{\epsilon}$ with $\epsilon\in(1/\sqrt{2},1)$ there will be a finite critical time $t_{\mathcal{B}}$ after which the two-spin state will become Bell local. Such ``sudden-death time'' is substantially postponed as the white noise becomes very small $(\epsilon\to 1)$, in which case Bell nonlocality, as measured by $\mathcal{B}(\rho_t^{\epsilon})$, will vanish only asymptotically [see Fig.~\ref{fig6}(a)]. For high levels of white noise ($\epsilon\leq1/\text{\small $\sqrt{2}$}$), Bell nonlocality never manifests itself.

\subsection{EPR steering}

EPR steering signalizes the capability of an observer to steer the state of a system in a remote site via local measurements~\cite{wiseman2007steering}. In scenarios where two measurements are performed per site on a two-qubit system, EPR steering becomes identical to Bell nonlocality, as demonstrated by Costa and Angelo in Ref.~\cite{costa2016quantification}. On the other hand, EPR steering and Bell nonlocality become distinguishable when at least three measurements are allowed per site, in which case the following EPR steering quantifier can be derived for the general two-qubit state \eqref{zeta}:
\be
\mathcal{S}(\zeta)=\max\left\{0,\text{\small$ \frac{\sqrt{\vec{c}\cdot\vec{c}}-1}{\sqrt{3}-1} $} \right\}.
\ee 
For the state under scrutiny here, this measure reduces to
\be
\mathcal{S}(\rho_t^{\epsilon})=\text{\small $\left(\frac{1+\sqrt{3}}{2}\right)$}\max \text{\small $\left\{0,\epsilon\sqrt{1+2\mathcal{E}_t^2}-1 \right\}$},
\label{St}
\ee
which indicates the existence of EPR steering as long as
\be
t<\sigma_0\text{\small $\sqrt{\ln\left(\frac{2\epsilon^2}{1-\epsilon^2}\right)}$}\equiv t_{\mathcal{S}}.
\label{tS}
\ee 
It follows that a sudden-death time $t_{\mathcal{S}}$ will exist for EPR steering whenever $\epsilon\in(1/\text{\small$\sqrt{3}$},1)$ and that $t_{\mathcal{S}}$ can be made arbitrarily large for reduced amounts of white noise [see Fig.~\ref{fig6}(a) for an illustration of this behavior]. For $\epsilon\leq1/\text{\small$ \sqrt{3}$}$, $\rho_t^{\epsilon}$ is nonsteerable\footnote{A method based on local hidden states and semidefinite program has been developed that predicts the presence of steering for $\epsilon > 1/2$~\cite{hirsch2016}, which agrees with the results of Ref.~\cite{wiseman2007steering}. The method employed here is based on linear steering inequalities (see Ref.~\cite{costa2016quantification} and references therein) and, although less powerful, allows for analytical analysis and is in full agreement with a recently introduced geometrical quantifier~\cite{ku2018}.}. 

\subsection{Entanglement}

Related to the degree of inseparability of a quantum state, entanglement can be computed for a two-qubit state $\rho$ by means of the concurrence~\cite{wootters1998entanglement}:
\be
E(\rho)\coloneqq\max\left\{0,\sqrt{\lambda_1}-\sqrt{\lambda_2}-\sqrt{\lambda_3}-\sqrt{\lambda_4} \right\},
\ee 
where $\lambda_1\geq\lambda_2\geq\lambda_3\geq\lambda_4$ are the eingenvalues of the operator $\rho(\sigma_y\otimes\sigma_y)\rho^*(\sigma_y\otimes\sigma_y)$ and $\rho^*$ is the complex conjugate of $\rho$. A straightforward calculation gives
\be
E(\rho_t^{\epsilon})=\tfrac{1}{2}\max\Big\{0,\epsilon(1+ 2\mathcal{E}_t)-1\Big\},
\label{Et}
\ee 
which predicts entanglement for 
\be
t<\sigma_0\text{\small $\sqrt{2\ln\left( \frac{2\epsilon}{1-\epsilon}\right)}$}\equiv t_E.
\label{tE}
\ee
A well defined instant $t_E$ will exist for entanglement sudden death if $\epsilon\in(1/3,1)$. While entanglement will vanish only asymptotically for reduced values of white noise ($\epsilon\to 1$), as illustrated in Fig.~\ref{fig6}(a), it will not occur if $\epsilon\leq 1/3$.

Two points are now worth noticing. First, in the regime of no white noise $(\epsilon=1)$, we have $E(\rho_t^{\epsilon=1})=\mathcal{E}_t$, which attaches an interesting interpretation to the damping factor. Moreover, we can revisit Section~\ref{sec3} and write the complementarity relation 
\be 
C_{\text{\tiny GME}}^2(\ket{\Psi_t})+E(\rho_t^{\epsilon=1})=1,
\label{complementarity}
\ee 
which explicitly shows that fourpartite entanglement develops over time at the expense of the two-spin entanglement. Second, in the domain $\epsilon\in(1/\text{\small$\sqrt{2}$},1)$, wherein the sudden-death times are all well defined, one has
\be 
t_{\mathcal{B}}<t_{\mathcal{S}}<t_{E},
\label{chronology}
\ee 
which corroborates the current knowledge according to which Bell nonlocality is the most fragile quantum resource, whereas entanglement is the least one~\cite{wiseman2007steering,costa2016quantification,costa2016generalized}.

\subsection{Quantum discord}

Introduced by Olliver and Zurek \cite{olliver2001quantum}, and independently by Henderson and Vedral \cite{henderson2001classical}, quantum discord was conceived as the difference between two different ways of quantifying mutual information for quantum states. Later on, Rulli and Sarandy \cite{rulli2011global} showed that quantum discord can also be viewed as the sensitivity of mutual information to minimally disturbing projective measurements conducted locally, i.e., in either of the particles. Using their formulation, the quantum discord of a bipartite state $\rho$ on $\mathcal{H_A\otimes H_B}$ is given by 
\be 
\mathcal{D_B}(\rho):=\min\limits_B\Big[ I(\rho)-I(\Phi_B(\rho)) \Big],
\ee
where $I(\rho)=S(\rho_{\mathcal{A}})+S(\rho_{\mathcal{B}})-S(\rho)$ is the mutual information, $S(\rho)=-\Tr(\rho\ln\rho)$ is the von Neumann entropy of $\rho$, and $\Phi_B(\rho)=\sum_b \left( \mathbbm{1}_{\mathcal{A}} \otimes B_b \right) \rho \left( \mathbbm{1}_{\mathcal{A}} \otimes B_b\right)$ is the state after a measurement of the observable $B=\sum_b bB_b$, with projectors $B_b=\ket{b}\bra{b}$ acting on $\mathcal{H}_{\mathcal{B}}$. To compute quantum discord in our model, we consider the observable $B=\hat{v}_2\cdot\vec{\sigma}$, with unit vector $\hat{v}_2(\theta_2,\phi_2)=(\cos\theta_2\sin\phi_2,\sin\theta_2\sin\phi_2,\cos\phi_2)$ and projectors $B_{\pm}=(\mathbbm{1}\pm\hat{v}_2\cdot\vec{\sigma})/2$. Direct calculations produce $S(\Tr_{S_{1,2}}\rho_t^{\epsilon})=S(\Phi_B(\Tr_{S_1}\rho_t^{\epsilon}))=\ln2$ and the formal result $\mathcal{D}_{S_2}(\rho_t^{\epsilon})=\min_B[S(\Phi_B(\rho_t^{\epsilon}))-S(\rho_t^{\epsilon})]$. Some more algebra gives
\be 
S(\rho_t^{\epsilon})=\text{\small $\left(\frac{1-\epsilon}{2}\right)\ln{2}+\left(\frac{1+\epsilon}{2}\right)H\left(\frac{1}{2}+\frac{\epsilon\,\mathcal{E}_t}{1+\epsilon}\right)+H\left(\frac{1+\epsilon}{2}\right)$},
\label{Srho}
\ee 
with the Shannon entropy $H(u)=-u\ln{u}-(1-u)\ln{(1-u)}$. Numerical analyses revealed that $(\theta_2,\phi_2)=(0,\tfrac{\pi}{4})$ define the optimal observable $B$ for all times, with which we have been able to compute $\min_BS(\Phi_B(\rho_t^{\epsilon}))$ and then obtain
\be
\mathcal{D}_{S_2}(\rho_t^{\epsilon})=\text{\small $\frac{1+\epsilon}{2}\left[\ln{2}- H\left(\frac{1}{2}+\frac{\epsilon\,\mathcal{E}_t}{1+\epsilon}\right)\right]$}.
\label{DS2t}
\ee 
In contrast with what we have for Bell nonlocality, EPR steering, and entanglement, there is no domain of $\epsilon$ for which the quantum discord of $\rho_t^{\epsilon}$ suddenly vanishes. In fact, regardless of the white-noise level, quantum discord vanishes only asymptotically (see Fig.~\ref{fig6}). By symmetry, one can straightforwardly conclude that $\mathcal{D}_{S_1}(\rho_t^{\epsilon})=\mathcal{D}_{S_2}(\rho_t^{\epsilon})$.

We can also quantify the sensitivity of total correlations to unread measurements conducted separately in both sites. This information is captured by the so-called symmetrical quantum discord~\cite{rulli2011global}, which for a state $\rho$ is formally written as
\be
\mathcal{D}(\rho):=\min\limits_{A,B}\big[I(\rho)-I(\Phi_{AB}(\rho))\big],
\ee
where $\Phi_{AB}(\rho)=\sum_{a,b}\left(A_a\otimes B_b \right)\rho\left(A_a\otimes B_b \right)$, for observables $A=\sum_aaA_a$ and $B=\sum_bbB_b$ acting on $\mathcal{H}_{S_1}$ and $\mathcal{H}_{S_2}$, respectively. Also in this case we have been able to analytically conduct all the calculations and prove that $\mathcal{D}(\rho_t^{\epsilon})=\mathcal{D}_{S_{1,2}}(\rho_t^{\epsilon})$. Hence, hereafter we make no distinction between quantum discord and its symmetrical counterpart.

\subsection{Irreality and realism-based nonlocality}

Now we discuss aspects of quantum irreality (in opposition to classical realism) by means of the framework put forward by Bilobran and Angelo~\cite{bilobran2015measure}. We start by looking at a quantifier of irreality---a measure that indicates by how much the hypothesis of realism is violated. From the premise that an element of reality will exist for $A$ after a measurement of this observable is realized for a given preparation $\rho$, these authors propose to take
\be
\mathfrak{I}_A(\rho):=S(\Phi_A(\rho))-S(\rho),
\label{irreality}
\ee
as a quantifier for the degree of irreality of $A$. This gives the entropic amount by which the preparation differs from a state of reality $\Phi_A(\rho)$. Clearly, if the preparation is already a state of reality for $A$, then $\mathfrak{I}_A(\Phi_A(\rho))=0$, and the classical notion of realism applies. Interestingly, for any $\rho$ on $\mathcal{H_A\otimes H_B}$, one shows that  $\mathfrak{I}_A(\rho)=\mathfrak{I}_A(\rho_{\mathcal{A}})+D_{A}(\rho)$, where $D_A(\rho)=I(\rho)-I(\Phi_A(\rho))$ is the measurement-dependent discord. [Note that $\mathcal{D_A}(\rho)=\min_AD_A(\rho)$.] This decomposition reveals that irreality actually captures both (i) information about local coherence and (ii) correlation changes induced by local measurements. In particular, for the state under scrutiny, because $\varrho_{S_1}=\Tr_{S_2}\rho_t^{\epsilon}=\mathbbm{1}/2$ it follows that $\Phi_A(\varrho_{S_1})=\varrho_{S_1}$ for all $A$ on $\mathcal{H}_{S_1}$, which implies that $\mathfrak{I}_A(\varrho_{S_1})=0$. As a consequence, $\mathfrak{I}_A(\rho_t^{\epsilon})=D_{S_1}(\rho_t^{\epsilon})$ and, therefore,
\be 
\mathfrak{I}_A(\rho_t^{\epsilon})\geq \mathcal{D}(\rho_t^{\epsilon}). 
\label{Irr>D}
\ee
This relation is important because it establishes a lower bound for the irreality of all observables $A$ on $\mathcal{H}_{S_1}$. Since for $\epsilon>0$ quantum discord vanishes only asymptotically, then it is guaranteed that no element of reality will exist at short times.  On the other hand, for the regime of maximum white noise ($\epsilon=0$), every observable will always be an element of reality, since $\rho_t^{\epsilon=0}=\mathbbm{1}/4$ and then $\mathfrak{I}_A(\rho_t^{\epsilon=0})=\mathcal{D}(\rho_t^{\epsilon=0})=0$.

\begin{figure}[htb]
\centering
\includegraphics[width=\linewidth]{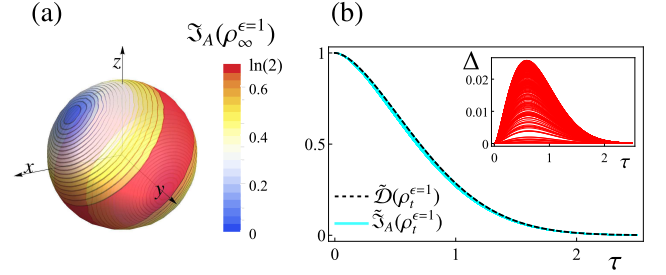}
\caption{(a) Contour plot for the asymptotic irreality $\mathfrak{I}_A(\rho_{\infty}^{\epsilon=1})$ of an observable $A=\hat{v}_1\cdot\vec{\sigma}$ on $\mathcal{H}_{S_1}$, in spherical coordinates, for the no-noise regime $(\epsilon=1)$. The color scale goes from zero (blue) to $\ln2$ (red). (b)~Scaled irreality $\tilde{\mathfrak{I}}_A(\rho_t^{\epsilon=1})$ (cyan curves) as a function of the scaled time $\tau=t/\sigma_0$ for 200 randomly chosen measurement directions $\hat{v}_1$. The scaled discord $\tilde{\mathcal{D}}(\rho_t^{\epsilon=1})$ (dashed black curve) defines a tight upper bound [see Eq.~\eqref{unification}]. In the inset, the difference $\Delta:=\tilde{\mathcal{D}}(\rho_t^{\epsilon=1})-\tilde{\mathfrak{I}}_A(\rho_t^{\epsilon=1})$ is plotted for each of the 200 measurement directions, showing results never greater than 0.03.}
\label{fig4}
\end{figure}

It is interesting to look also at the no-noise regime. Introducing the unit vector $\hat{v}_1(\theta_1,\phi_1)$ to define a generic observable $A=\hat{v}_1\cdot\vec{\sigma}$ for the spin $S_1$, we find a lengthy and nonenlightening analytical function for $\mathfrak{I}_A(\rho_t^{\epsilon})$ (omitted). For $\epsilon=1$, though, an interesting universal behavior is found. Since in this case the initial state (the singlet) is rotationally invariant, any observable is maximally unreal at $t=0$. As the Gaussian packets start to split themselves and get correlated with the spins, irreality becomes direction dependent and typically decays with time, eventually reaching the asymptotic value
\be
\mathfrak{I}_A(\rho_{\infty}^{\epsilon=1})=H\left(\tfrac{1+\nu_{\theta_1\phi_1}}{2}\right),
\label{irreality_limit}
\ee
where $\nu_{\theta\phi}=(\cos{\phi}+\cos{\theta}\sin{\phi})/\text{\small$\sqrt{2}$}$. A panoramic view of the asymptotic irreality is presented in Fig.~\ref{fig4}(a). First of all, it is seen that $\mathfrak{I}_A(\rho_{\infty}^{\epsilon=1})=0$ only for two particular observables, namely, $\pm(\sigma_x+\sigma_z)/\text{\small $\sqrt{2}$}$ (center of the blue circle and its antipode), which are directly related to the quantum coin \eqref{coin} that we have adopted for the walk. This happens because the position of each walker correlates with its respective coin, thus establishing its reality. For any other observable, we have $\mathfrak{I}_A(\rho_{\infty}^{\epsilon=1})>0$, which reveals a broad scenario of quantum irreality. In particular, since the Shannon entropy $H(u)$ reaches its maximum for $u=1/2$, there is a continuous set of observables, defined by $\nu_{\theta_1\phi_1}=0$, for which the asymptotic irreality reaches the maximum value $\ln{2}$. This set corresponds to the center of the red strip. We have checked that, in fact, these observables remain maximally unreal for every instant of time. 

Interestingly, we also found that the way irreality gets to the asymptote~\eqref{irreality_limit} is nearly direction-independent (as long as we exclude the aforementioned maximal-irreality set). After some numerical incursions, we have been able to show that 
\be 
\tilde{\mathfrak{I}}_A(\rho_t^{\epsilon=1}):=\text{\small $\frac{\mathfrak{I}_A(\rho_t^{\epsilon=1})-\mathfrak{I}_A(\rho_{\infty}^{\epsilon=1})}{\mathfrak{I}_A(\rho_0^{\epsilon=1})-\mathfrak{I}_A(\rho_{\infty}^{\epsilon=1})}\lessapprox \frac{\mathcal{D}(\rho_t^{\epsilon=1})}{\mathcal{D}(\rho_0^{\epsilon=1})}$}=:\tilde{\mathcal{D}}(\rho_t^{\epsilon=1}),
\label{unification}
\ee 
with $\mathfrak{I}_A(\rho_0^{\epsilon=1})=\mathcal{D}(\rho_0^{\epsilon=1})=\ln{2}$. The cyan curves presented in Fig.~\ref{fig4}(b) illustrate the behavior of the scaled irreality $\tilde{\mathfrak{I}}_A(\rho_t^{\epsilon=1})$ for 200 randomly chosen directions $\hat{v}_1(\theta_1,\phi_1)$ as a function of the scaled time $\tau=t/\sigma_0$. Clearly, the curves do not significantly deviate from each other and are all upper bounded by the scaled discord $\tilde{\mathcal{D}}(\rho_t^{\epsilon=1})$ (black dashed line). As shown in the inset, $0\leq \tilde{\mathcal{D}}(\rho_t^{\epsilon=1})-\tilde{\mathfrak{I}}_A(\rho_t^{\epsilon=1})<0.03$. Hence, to a pretty good accuracy we can state that the scaled irreality is determined by the scaled discord, which is observable independent. It follows, therefore, that there is an approximate class of universality for the irreality behavior, which is likely to emerge from the fact that the initial state of the spins is the rotationally invariant singlet.

\begin{figure}[ht]
\centering
\includegraphics[scale=0.55]{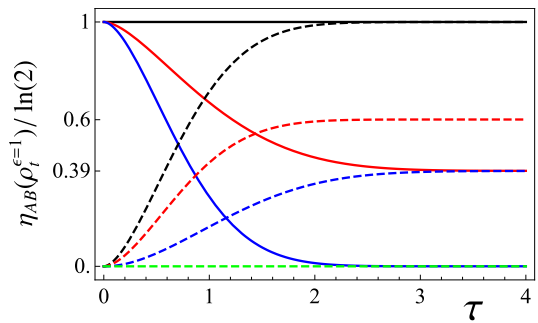}
\caption{Behavior of the normalized contextual realism-based nonlocality $\eta_{AB}(\rho_t^{\epsilon=1})/\ln{2}$, in the no-noise regime ($\epsilon=1$), as a function of the scaled time $\tau=t/\sigma_0$, for observables $A$ and $B$ (contexts) obeying the code that follows. In continuous lines, from top to bottom: $A=B=\sigma_y$ (black line), $A=B=\sigma_z$ (red line), $A=B=(\sigma_x+\sigma_z)/\text{\scriptsize $\sqrt{2}$}$ (blue line); in dashed lines, from top to bottom: $A=(\sigma_x-\sigma_z)/\text{\scriptsize $\sqrt{2}$}$ and $B=\sigma_y$ (black line), $A=\sigma_x$ and $B=\sigma_y$ (red line), $A=\sigma_z$ and $B=\sigma_x$ (blue line), $A=(\sigma_x+\sigma_z)/\text{\scriptsize $\sqrt{2}$}$ and $B=\sigma_y$ (green line).}
\label{fig5}
\end{figure}

It is clear from the above that, in contrast with all the other types of quantumness studied so far, irreality can be preserved during the quantum walk. Presumably, a similar behavior can exist for the realism-based nonlocality, a notion that has shown to be dramatically different from Bell nonlocality~\cite{bilobran2015measure}. In its contextual version, it is defined as
\be
\eta_{AB}(\rho):=\mathfrak{I}_A(\rho)-\mathfrak{I}_A(\Phi_B(\rho)),
\ee 
for $\rho$ on $\mathcal{H_A\otimes H_B}$. By construction, this measure captures alterations in the irreality of $A$ induced by measurements of $B$ conducted in a far remote site $\mathcal{B}$. Using its symmetrical expansion $\eta_{AB}(\rho)=S(\Phi_A(\rho))+S(\Phi_B(\rho))-S(\Phi_{AB}(\rho))-S(\rho)$, with $\Phi_A(\Phi_B(\rho))=\Phi_B(\Phi_A(\rho))\equiv \Phi_{AB}(\rho)$, it can be verified that irreality is a necessary condition for the existence of this type of nonlocality, since $\eta_{AB}(\Phi_{A}(\rho))=\eta_{AB}(\Phi_{B}(\rho))=0$. Following the above formulation, we can compute the contextual realism-based nonlocality $\eta_{AB}(\rho_t^{\epsilon})$ for the context defined by generic observables $A=\hat{v}_1\cdot\vec{\sigma}$ and $B=\hat{v}_2\cdot\vec{\sigma}$. For the maximum-noise scenario, we directly obtain $\eta_{AB}(\rho_t^{\epsilon=0})=0$, since for the state $\rho_t^{\epsilon=0}=\mathbbm{1}/4$ all observables are elements of reality. On the other hand, in the other extreme $(\epsilon=1)$, all sorts of behaviors can be found for the contextual realism-based nonlocality, as is illustrated in Fig.~\ref{fig5}. For the asymptotic values of the contextual realism-based nonlocality we have found
\be 
\eta_{AB}(\rho_{\infty}^{\epsilon=1})=H\left(\tfrac{1+\nu_{\theta_1\phi_1}}{2}\right)+H\left(\tfrac{1+\nu_{\theta_2\phi_2}}{2}\right)-H\left(\tfrac{1+\nu_{\theta_1\phi_1}\nu_{\theta_2\phi_2}}{2}\right).
\ee 
Therefore, there exists an infinite set of observables, defined by $(\nu_{\theta_1\phi_1},\nu_{\theta_2\phi_2})=(0,0)$, for which the contextual realism-based nonlocality will asymptotically reach its maximum value $\ln{2}$. 

\begin{figure}[htb]
\centering
\includegraphics[width=\linewidth]{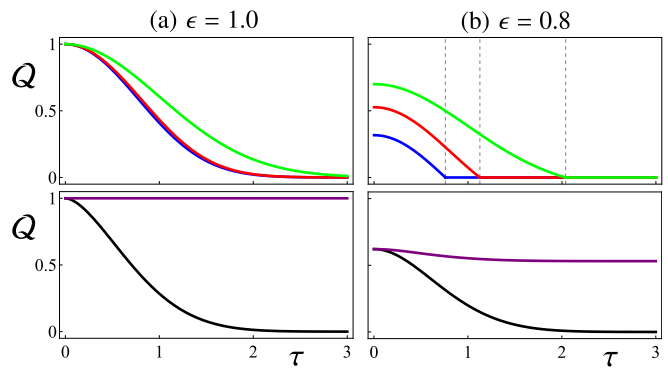}
\caption{All the (observable independent) quantumness quantifiers $\mathcal{Q}$ computed in this work for the two-spin state $\rho_t^{\epsilon}$ as a function of the scaled times $\tau=t/\sigma_0$ for (a) $\epsilon=1.0$ (left panels) and (b) $\epsilon=0.8$ (right panels). $\mathcal{Q}$ assumes, in upper panels, Bell nonolocality $\mathcal{B}$ (blue lower line), EPR steering $\mathcal{S}$ (red middle line), and entanglement $E$ (green upper line), and, in bottom panels, normalized (symmetrical) quantum discord $\mathcal{D}/\ln{2}$ (black lower line), and normalized realism-based nonlocality $\mathcal{N}/\ln{2}$ (purple upper line). The vertical dashed lines in the upper right panel refer to the sudden-death times given by Eqs.~\eqref{tB}, \eqref{tS}, and \eqref{tE}. Quantumness typically decreases with both time and the amount $1-\epsilon$ of noise, but realism-based nonlocality survives.}
\label{fig6}
\end{figure}

It is also interesting to look at the realism-based nonlocality~\cite{gomes2018nonanomalous}, which makes reference solely to the quantum state, thus having no link whatsoever with particular contexts:
\be
\mathcal{N}(\rho):=\max_{A,B} \eta_{AB}(\rho).
\label{rbn}
\ee 
Besides being a sufficient condition for the existence of contextual realism-based nonlocality, it has been proved that $\mathcal{N}$ is nonanoumalous~\cite{gomes2018nonanomalous} and nonincreasing under the action of local maps~\cite{gomes2019resilience}. Even though the maximization over $\{A,B\}$ implies a hard mathematical problem in general, numerical and analytical incursions on $\eta_{AB}(\rho_t^{\epsilon})$ give us the clues for the accomplishment of such a task. For instance, we see from Fig.~\ref{fig5}, that the choice $A=B=\sigma_y$ is optimal. In fact, we have verified that parallel direction measurements $\hat{v}(\theta,\phi)$ satisfying $\nu_{\theta\phi}=0$, that is, observables in the circle represented by the red strip in Fig.~\ref{fig4}(a), provide the maximization. We then find
\be
\mathcal{N}(\rho_t^{\epsilon})=\mathcal{D}(\rho_t^{\epsilon})+H\text{\small $\left(\frac{1+\epsilon\,\mathcal{E}_t}{2}\right)$}-H\text{\small $\left(\frac{1+\epsilon}{2}\right)$},
\label{Nt}
\ee 
with the symmetrical quantum discord $\mathcal{D}(\rho_t^{\epsilon})$ being given by Eq.~\eqref{DS2t}. Realism-based nonlocality is similar to quantum discord in that they never experiment sudden death. On the other hand, while the latter vanishes asymptotically, the former behaves as $\mathcal{N}(\rho_{\infty}^{\epsilon})=\ln{2}-H\left(\tfrac{1+\epsilon}{2}\right)$ (see Fig.~\ref{fig6} for an illustration), which vanishes as $t\to\infty$ only in the maximum noise regime ($\epsilon=0$). In fact, it directly follows from Eq.~\eqref{Nt} that $\mathcal{N}(\rho_t^{\epsilon})\geq \mathcal{D}(\rho_t^{\epsilon})$. Therefore, in flagrant contrast with the other nonclassical features, $\mathcal{N}(\rho_t^{\epsilon})$ manifests itself as the most resilient one, which is in full agreement with previously conducted studies~\cite{gomes2019resilience}.

\section{Concluding remarks}
\label{sec5}

Quantum walk studies often demand numerical simulations, which do not always allow for the access of some refined physical aspects. In this work, by introducing a Gaussian model, which proved to be quite accurate for delocalized walkers $(\sigma_0\gg 1)$, we were able to conduct a profound analysis of the nonclassical features dynamics in a two-walker system. Previous studies~\cite{valcarcel2010tailoring,romanelli2010distribution,romanelli2012thermodynamic,romanelli2014entanglement,zhang2016creating,orthey2017asymptotic,orthey2019connecting,ghizoni2019trojan} allow us to optimistically speculate upon the applicability of our model even to scenarios involving more localized states.

Starting with a single quantum walker, we derived an analytical expression for the entanglement between spin and position [Eq.~\eqref{ESX}]. This result reveals that the production of entanglement is regulated by a parameter that controls the initial coherence of the spin state. For the problem of two noninteracting walkers, with spins prepared in the singlet state, we showed that genuine fourpartite entanglement is created throughout the walk, monotonically increasing with time [Eq.~\eqref{4Et}], at the expense of two-spin correlations [Eq.~\eqref{complementarity}]. This reveals a scenario where the total amount of resource is conserved. Also, we found that by measuring the sign of the relative coordinate, the spins can be prepared in superpositions of Bell states.

With respect to nonclassical aspects between the spins, our results are graphically summarized in Fig.~\ref{fig6}, for two noise regimes, where the quantifiers are separated into two rows, according to their susceptibility to sudden death. The panels in the upper row show the behaviors over time of Bell nonlocality~[Eq.~\eqref{Bt}], EPR steering~[Eq.~\eqref{St}], and entanglement~[Eq.~\eqref{Et}], while in the lower row simulations are presented for (symmetrical) quantum discord~[Eq.~\eqref{DS2t}] and realism-based nonlocality~[Eq.~\eqref{Nt}]. Besides showing a clear chronology of deaths, which is formally stated in the relations~\eqref{chronology}, our findings corroborate the view according to which there is a strict hierarchy~\cite{gomes2019resilience} among the quantifiers, in such a way that the existence of Bell nonlocality implies steering, which implies entanglement, which implies quantum discord, which then implies realism-based nonlocality, while the converse sequence of implications is false. Moreover, it is clear that realism-based nonlocality is the only type of quantumness that survives upon the noisy channels considered. From such aspect, an urgent demand arises aiming at characterizing the potential of this quantumness as a useful quantum resource. 

Finally, from a foundational viewpoint, lessons can be learned with respect to (ir)reality. According to the relation~\eqref{Irr>D}, since quantum discord vanishes only asymptotically, generic spin variables cannot be elements of reality. In fact, given the presence of fourpartite quantum correlations, the positions cannot be either. There are only two specific spin observables that asymptotically behave as elements of reality, and these are closely related with the quantum coin. Altogether, our findings reinforce the potential of quantum walks as a rich arena for studies involving information-theoretic and foundational issues, such as the interconversion of bipartite to multipartite entanglement and the dynamics of further nonclassical aspects, from nonlocality to (multipartite) quantum correlations and violations of classical realism. 

\begin{acknowledgments}
This study was financed in part by the Coordena\c{c}\~ao de Aperfei\c{c}oamento de Pessoal de N\'ivel Superior, Brasil (CAPES), Finance Code 001. A.C.O. thanks Ana C. S. Costa for insightful suggestions and Centro de Assessoria de Publica\c{c}\~ao Acad\^{e}mica (CAPA) for writing assistance. R.M.A. acknowledges support from CNPq/Brazil (Grant No. 303111/2017-8) and the National Institute for Science and Technology of Quantum Information (CNPq, Grant No. INCT-IQ 465469/2014-0).
\end{acknowledgments}


\end{document}